\documentclass[conference]{IEEEtran}
\IEEEoverridecommandlockouts
\usepackage{amsmath,amsfonts,amsthm}
\usepackage{algorithmic}
\usepackage{algorithm}
\usepackage{array}
\usepackage[caption=false,font=normalsize,labelfont=sf,textfont=sf]{subfig}
\usepackage{textcomp}
\usepackage{stfloats}
\usepackage{url}
\usepackage{verbatim}
\usepackage{graphicx}
\usepackage[acronym]{glossaries}
\usepackage{todonotes}
\usepackage{cite}
\usepackage{hyperref}   
\usepackage{booktabs}
\usepackage{tabularx}
\usepackage{threeparttable}
\usepackage{listings}
\usepackage[frozencache,cachedir=.]{minted}
\usepackage{textcomp}
\usepackage{xcolor}

\def\BibTeX{{\rm B\kern-.05em{\sc i\kern-.025em b}\kern-.08em
    T\kern-.1667em\lower.7ex\hbox{E}\kern-.125emX}}

\theoremstyle{definition}
\newtheorem{prob}{Problem}
\newtheorem{sol}{Solution}

\makeatletter
\newcommand{\newlineauthors}{%
  \end{@IEEEauthorhalign}\hfill\mbox{}\par
  \mbox{}\hfill\begin{@IEEEauthorhalign}
}
\makeatother

\newacronym{uq}{UQ}{Uncertainty Quantification}
\newacronym{hpc}{HPC}{High-Performance Computing}
\newacronym{gp}{GP}{Gaussian Process}
\newacronym{slr}{SLR}{Schedule Length Ratio}
\newacronym{http}{HTTP}{Hypertext Transfer Protocol}
\newacronym{umbridge}{UM-Bridge}{the UQ and Modelling Bridge}
\newacronym{hq}{HQ}{HyperQueue}
\newacronym{mpi}{MPI}{Message Passing Interface}

\makeglossaries

\begin{document}
%TODO too long? @Mikkel, James, Will: Should we mention the GS2 application in the title?
%\title{A Performance Analysis of High-Volume, Low-Complexity Task Scheduling on HPC Systems}
\title{A Performance Analysis of Task Scheduling for UQ Workflows on HPC Systems}

%\title{Efficient and User-friendly Task Scheduling for UQ Applications on HPC Systems}

%No particular thought to author order yet
\author{\IEEEauthorblockN{Chung Ming Loi, Anne Reinarz}
\IEEEauthorblockA{\textit{Department of Computer Science} \\
\textit{Durham University}\\
Durham, UK \\
\{chung.m.loi,anne.k.reinarz\}@durham.ac.uk}
\and
 \IEEEauthorblockN{Linus Seelinger}
 \IEEEauthorblockA{\textit{Scientific Computing Center} \\
 \textit{Karlsruhe Institute of Technology}\\
 Karlsruhe, Germany \\
linus.seelinger@kit.edu}
\newlineauthors
\IEEEauthorblockN{William Hornsby, James Buchanan}
\IEEEauthorblockA{\textit{UKAEA-CCFE} \\
Abingdon, UK \\
\{william.hornsby, james.buchanan\}@ukaea.uk}
\and
\IEEEauthorblockN{Mikkel Lykkegaard}
\IEEEauthorblockA{\textit{Danish Technological Institute} \\
Aarhus, Denmark \\
mbly@teknologisk.dk}
}
 % \IEEEauthorblockN{Darius Plesan-Tohoc}
 % \and


\maketitle
\begin{abstract} 

\acrfull{uq} workloads are becoming increasingly common in science and engineering. They involve the submission of thousands or even millions of similar tasks with potentially unpredictable runtimes, where the total number is usually not known a priori. A static one-size-fits-all batch script would likely lead to suboptimal scheduling, and native schedulers installed on \acrfull{hpc} systems such as SLURM often struggle to efficiently handle such workloads. In this paper, we introduce a new load balancing approach suitable for \acrshort{uq} workflows. To demonstrate its efficiency in a real-world setting, we focus on the GS2 gyrokinetic plasma turbulence simulator. Individual simulations can be computationally demanding, with runtimes varying significantly—from minutes to hours—depending on the high-dimensional input parameters. Our approach uses \acrlong{umbridge}, which offers a language-agnostic interface to a simulation model, combined with \acrlong{hq} which works alongside the native scheduler. In particular, deploying this framework on \acrshort{hpc} systems does not require system-level changes. We benchmark our proposed framework against a standalone SLURM approach using GS2 and a \acrlong{gp} surrogate thereof. Our results demonstrate a reduction in scheduling overhead by up to three orders of magnitude and a maximum reduction of 38\% in CPU time for long-running simulations compared to the na\"ive SLURM approach, while making no assumptions about the job submission patterns inherent to UQ workflows.
\end{abstract}


\section{Introduction}
Efficient scheduling of applications is critical for achieving high performance on large compute clusters. Traditional resource and job management software such as SLURM \cite{SLURM}, PBS \cite{PBS} or LoadLeveler \cite{loadleveler} are designed to schedule large, long-running, homogeneous applications. Ahn et al. \cite{flux} note that on production clusters at Lawrence Livermore National Laboratory 48.1\% of jobs now involve submission of at least 100 identical jobs within a short time frame. In other words, workloads consisting of a large number of simple tasks are quickly becoming the norm. For these types of small tasks, traditional schedulers require users to manually balance their workloads, either by submitting large numbers of jobs quickly or by manually aggregating these tasks. Both approaches commonly lead to suboptimal resource utilisation.

An area in which these types of workload are particularly prevalent is \acrfull{uq}. Whether it is uncertainty propagation or Bayesian inference \cite{bay_inverse}, the same simulation is run thousands of times with different input parameters. This is true for a wide range of algorithms; to list just a few Latin hypercube, quasi-Monte Carlo, ADVI, HMC, and NUTS \cite{kucukelbir2017automatic, duane_hybrid_1987, hoffman2014no}. In this paper, we will focus \acrshort{uq} workloads enabled by the \acrfull{umbridge} framework \cite{joss, SEELINGER2024113542}, a universal interface that makes numerical models accessible from any programming language or higher-level software, with a main focus on \acrshort{uq}. It consists of an \acrfull{http} that can query the model for evaluations and several types of derivatives. 

As a realistic use-case, we apply our methods to simulations of plasma micro-turbulence. Plasma turbulence determines the heat confinement properties of fusion reactors.  Direct numerical calculation of the transport coefficients and the micro-instabilities that form the turbulence is computationally expensive and is a significant bottleneck in fusion plasma modelling \cite{instabilities}. The considerable number of geometric and thermodynamic parameters, the interactions that influence these coefficients and the resolutions needed to accurately resolve these modes, make direct numerical simulation for parameter space exploration computationally extremely challenging.  The runtime of the simulation varies strongly between a few minutes up to several hours on a single node with two AMD EPYC 7702 64-core processors. Furthermore, the simulation runtime is not easily predicted for a given set of inputs. 
Therefore, significant effort is being invested in the production of machine-learnt surrogate models built from datasets of thousands of simulations \cite{10.1063/5.0174478,Zanisi_2024,Narita_2019, shinya_2024, 10.1063/1.5134126, Rodriguez-Fernandez_2024}.

The end goal in our use case is the computation of multiple integrals for a given set of input parameters, as described in \cite{integral}. Such computations are common in \acrshort{uq} workflows, where objectives often include evaluating expectations of quantities of interest or determining the maximum a posteriori probability point. Efficient evaluation of these integrals requires numerous forward model evaluations at different integration points. These evaluations are not always embarrassingly parallel, e.g. due to dependencies between integration points, such as in adaptive quadrature methods. Additionally, significant variations in runtime across integration points complicate scheduling. 

On \acrfull{hpc} systems, one approach to this scheduling problem would be to incorporate both the physical (forward) model and the \acrshort{uq} algorithm into one monolithic implementation. This approach can be facilitated by, e.g. Dask \cite{dask} and Ray \cite{ray}. Both are Python libraries for parallel computing which can be embedded as meta-schedulers within a larger job managed by traditional \acrshort{hpc} schedulers. This allows for fine-grained scheduling of tasks but requires the user to manually handle resource allocation by manually creating new SLURM/PBS jobs in order to utilise more resources. This manual handling may lead to a worse user experience and resource underutilisation. Dask and Ray are specific to Python workloads, and do not work out of the box as a generalised solution for any tasks written in other languages. In many real world settings, the main bottleneck of Dask lies in its runtime overhead \cite{daskdisc}. Furthermore, these approaches require the \acrshort{uq} specialist and the application specialist to collaborate throughout the project. The \acrshort{uq} algorithm and modelling software need to be written in compatible languages and use compatible parallelisation techniques, furthering the complexity of incorporating such an approach. 

Alternatively, it is possible to use workflow management tools such as Snakemake \cite{snakemake}  that can aggregate multiple tasks into groups. This helps reduce the number of jobs submitted and thereby decreases the overhead introduced by \acrshort{hpc} schedulers such as SLURM/PBS. The main limitation is that Snakemake performs this grouping eagerly, i.e. prior to the execution of tasks. This means that tasks are not dynamically balanced or scheduled in real-time, leading to sub-optimal node and core utilisation.  Merlin \cite{merlin} also offers load balancing and automatic task aggregation. However, the load balancing capabilities is constrained since Merlin requires users to define task queues with fixed concurrency—meaning no dynamic resource allocation. This tool is also complicated to set up, as it has dependencies on multiple external services, making it less accessible for many potential users. In contrast, \acrfull{hq} \cite{hyperqueue} has minimal dependencies, it is deployable to a large number of systems and allow for dynamical load balancing across all allocated nodes and CPU cores. \acrshort{hq} does this by deploying a lightweight server that manages the task scheduling and distribution across the computing cluster. This is similar to Kubernetes, where it coordinates a network of worker nodes that execute these tasks. These worker nodes can be manually designated or spawned on demand with a \textit{worker allocation}—allowing for the scaling of compute resources with workload demands. This setup not only improves resource utilisation across the cluster, but also reduces the overhead and complexity associated with managing a large number of low-complexity tasks. Similar tools such as QCG-PilotJob have been proposed in the context of the VECMAtk project \cite{doi:10.1098/rsta.2020.0221} to integrate with the EasyVVUQ framework.

This paper will address the challenges associated with effectively distributing a large number of loosely-coupled tasks across existing \acrshort{hpc} environments. In order to simplify the shift towards better scheduling solutions, we investigate methods that can function atop existing \acrshort{hpc} schedulers without requiring wide system-level changes or administrative privileges. We also aim to minimise the technical expertise required, making utilising \acrshort{hpc} resources with low complexity tasks easier for application specialists, such as engineers or geophysicists. We use the \acrshort{umbridge} framework \cite{SEELINGER2024113542}, which has been designed to improve accessibility of state-of-the-art \acrshort{uq} methods. It originally provided configurations mainly for use on smaller, self-administered clusters and cloud-based scaling using Kubernetes. In this paper we extend its use to classical \acrshort{hpc} systems by introducing a new load balancer that integrates with SLURM.

We will first introduce the principles behind the \acrshort{umbridge} protocol and the existing cloud-native Kubernetes implementation as a direct comparison to the work of this paper. We will then demonstrate a custom load balancer implementation for \acrshort{umbridge} using \acrshort{hq}. To validate the proposed solution in a realistic setting, we provide a real-world case study involving evaluations of the gyrokinetic plasma turbulence simulator GS2, as well as a \acrfull{gp} surrogate of the same simulation. 

\section{Load Balancers and UM-Bridge} \label{load balancer}
 %UQ algorithms frequently require a simulation to be run repeatedly over a range of input parameters as a part of the sampling process. This can be achieved by intrusively changing the source code to ``randomise'' the deterministic model. However, it requires domain experts in both areas (UQ and physical model) to efficiently implement the code. The non-intrusive method has the advantage to separate concerns, which often leads to a simpler development process. In both cases, they usually create a large number of low complexity tasks that put strain on traditional HPC schedulers. Additionally if the models have varying runtimes it may be important to load balance the model evaluations. 

This section outlines the integration of \acrshort{umbridge} with applications, covering containerisation strategies and our new load balancing solution which utilises \acrshort{hq}.

%%% mention programming language agnostic
\subsection{UM-Bridge}
Many \acrshort{uq} algorithms consist of repeated evaluation of a so-called forward model with different input parameters. These forward models, which represent complex physical systems, can be computationally intensive. For example, the most demanding evaluation of the GS2 model in our experiments required approximately three hours of computation on 8 cores of AMD EPYC 7702 processor. However, this runtime applies to the linear mode of the GS2 simulation, which simplifies the system by assuming perturbations grow or decay without coupling to other modes. Nonlinear simulations, which capture the full dynamics of turbulence and include mode coupling, are orders of magnitude more computationally expensive. 

\begin{figure*}[h]
    \centering
    \includegraphics[width=0.9\textwidth]{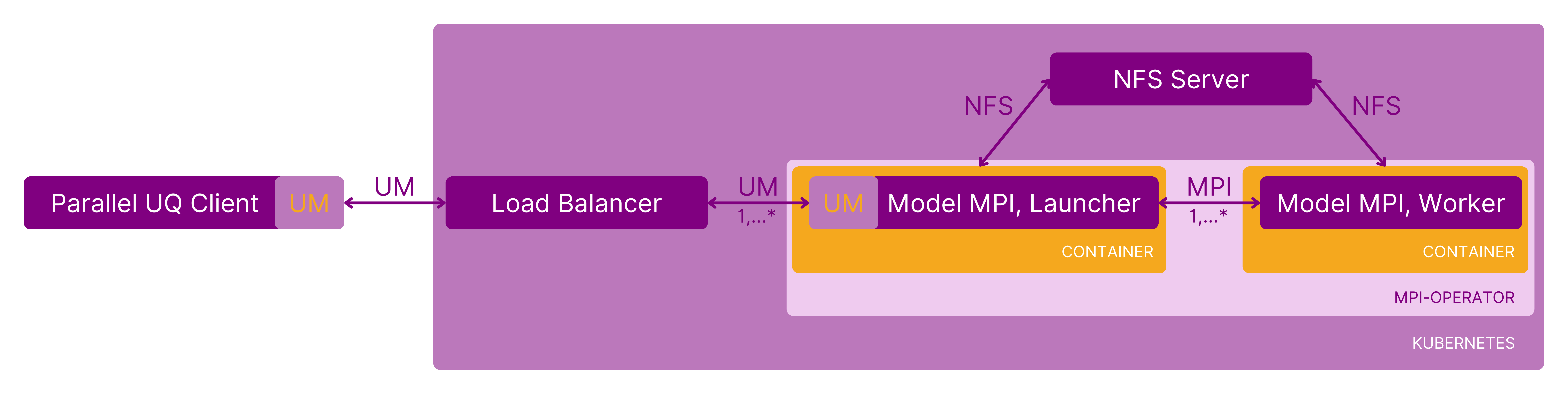}
    \includegraphics[width=0.9\textwidth]{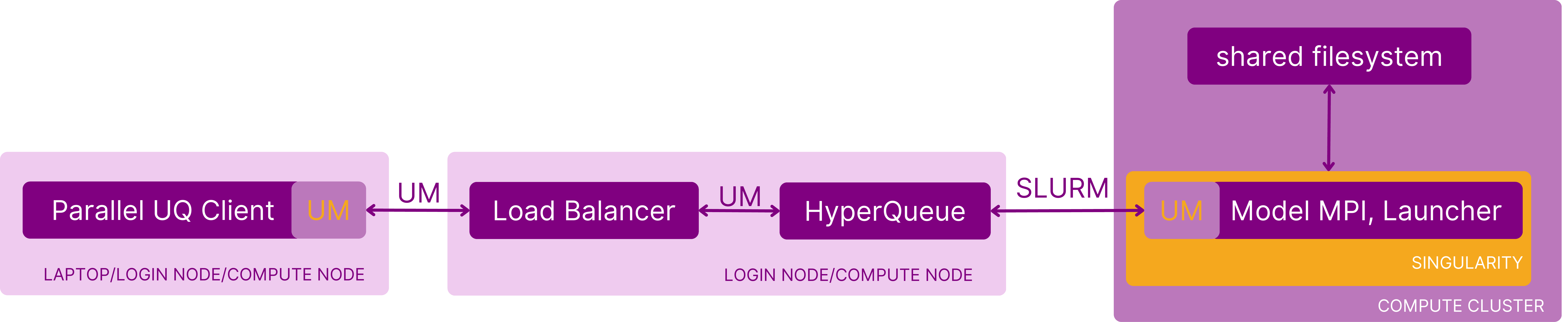}
    \caption{Top: Pre-defined Kubernetes configuration for parallel instances of any UM-Bridge model container. Bottom: Load balancer configuration for parallel instances of any UM-Bridge model container.}
    \label{fig:umbridge-setup}
\end{figure*}

\acrshort{umbridge} \cite{joss, SEELINGER2024113542} was introduced to link such forward models with \acrshort{uq} software non-intrusively. Through this link between UQ software and forward models, it is possible to introduce parallelisation to both the forward model and the \acrshort{uq} algorithm. The main aim is to provide non-experts a straightforward starting point for scaling up their applications.

In the abstract, we define a model simply as a map $$F:\mathbb{R}^n \rightarrow \mathbb{R}^m,$$ taking a parameter vector $\theta$ onto a model output $F(\theta)$. This map will be evaluated at a finite set of points $\{\theta\}_{i=0}^N$. These points may be known a-priori as in most propagation algorithms or in the construction of surrogate models; or depend on previous points as in the Bayesian inversion setting. Some \acrshort{uq} algorithms also require derivatives of $F$, usually in terms of a Jacobian or Hessian. 

The key idea behind \acrshort{umbridge} is to implement the above abstract mathematical interface directly in software. We treat \acrshort{uq} algorithms and numerical models as separate applications, linked only through an \acrshort{http} based network protocol as clients and servers. The role of the load balancer is to then distribute the evaluation requests $\{F(\theta_i)\}_{i=0}^N$ across the available nodes/CPU cores. Note that the \acrshort{uq} algorithm handles data dependencies and the load balancer only distributes the resulting requests. 

\subsection{Containerisation}
To facilitate reproducibility and system independence, \acrshort{umbridge} uses containerisation. This approach is well suited to cloud environments, where Kubernetes can orchestrate the containerised workloads \cite{googleblog}. In particular, the use of Kubernetes allows \acrshort{umbridge} users to dynamically scale computational resources as needed, which is particularly helpful for the variable workload demands of some \acrshort{uq} algorithms \cite{SCpaper}. The most prominent issue with this approach, however, is the cost, as this can easily skyrocket depending on resource demand and utilisation. For users who have access to alternative resources such as a \acrshort{hpc} cluster, it is important to be able to utilise these effectively.

%\todo{find ref for cloud-based scheduler}
%\todo{LS: Careful! k8s can be deployed on-prem! Separate arguments regarding public cloud vs. k8s. Also, public cloud can potentially be a cheaper option if you can't permanently utilise an on-prem HPC system.}
%In contrast to traditional on-premises HPC systems, a typical cloud computing environment consists of a large number of flexible, configurable compute nodes, each with a dynamic amount of resources that can be used. This provides a dynamic architecture that can allow for the scaling of resources in real-time, adapting to the computational demand of tasks which may fluctuate in both complexity and volume. This model has led to the development of cloud-based schedulers that excel at efficient resource utilisation for many types of applications. In order for this dynamic scaling to be made possible, underlying server configurations and application deployments require portability, which led to the popularisation of containerisation technologies such as Docker and Singularity.

%%% stuff about system admin can be removed (Done)
In classical \acrshort{hpc} systems, containerisation remains valuable, as it provides portability and reproducibility by simplifying deployment across systems. \acrshort{hpc} applications often come with complex software dependencies, sometimes not all required software dependencies are installed on a given cluster. Even when users install software locally, such as by building from source, they may encounter conflicts due to incompatible software versions. Containers address these challenges by allowing users to package all necessary dependencies along with the application into a self-contained environment. This eliminates the need for manual installation on the cluster. However, achieving optimal integration often requires adapting container orchestration to align with the scheduling constraints and resource management policies of traditional \acrshort{hpc} systems, such as SLURM or PBS.

While Docker stands out as the \textit{de facto} standard of containers, it is widely considered unsuitable for \acrshort{hpc} usage due to security concerns \cite{docker-security}. Furthermore, \acrshort{hpc} focuses on application performance, this means containers must exploit the underlying hardware efficiently to match the system utilisation of non-containerised applications. Numerous studies indicate Singularity containers achieved near baremetal efficiency. This makes Singularity a good candidate for containerised workloads in \acrshort{hpc}\cite{10.1007/s10586-021-03460-8, 8950978}.%\todo[inline]{We further demonstrate Singularity's scaling performance in Figure  using a moderately compute intensive GS2 workload. What happened to this?}

The Kubernetes setup is shown at the top of Figure \ref{fig:umbridge-setup}. \acrshort{uq} clients may send multiple concurrent evaluation requests to the cluster. Through model side load balancing and the abstraction \acrshort{umbridge} provides, the parallelisation strategies of \acrshort{uq} and model codes are fully separated, and thread parallelism or asynchronous code will typically suffice on the \acrshort{uq} side \cite{SEELINGER2024113542}. \acrfull{mpi} parallelism across containers is also supported by this Kubernetes configuration, albeit with some minor assumptions on how the container images are constructed. This provides \acrshort{umbridge} containers with full \acrshort{mpi} support and an additional shared file system while running in a (containerised) standard software environment. A performance study in \cite{SEELINGER2024113542} has shown the Kuberenetes setup scales well up to 5600 virtual CPUS.

%Existing parallel software is therefore easy to incorporate. Resulting performance rivals that of traditional HPC systems while avoiding the typically very time consuming machine specific software setup.

\subsection{SLURM and HyperQueue Integration}
%%% Paraphrase the Kuberenets costs etc paragraph in intro here
%\todo[inline]{LS: k8s cost is not a solid argument here, see note above}
%Although Kubernetes is a suitable candidate for heterogeneous workloads, the cost with its service is not sustainable for the purpose of academic research. Nevertheless, the flexibility of Kubernetes remains desirable to users who have access to alternative resources like a HPC cluster.

\acrshort{hq} was designed to handle the submission of many small jobs \cite{hyperqueue}, similar to the job array function in SLURM. However, \acrshort{hq} offers more flexibility in the specification of resources among jobs, which helps with scheduling efficiency and usability. In order to use \acrshort{hq} for \acrshort{uq} workflows, we have written an interface that passes \acrshort{umbridge} client requests to an \acrshort{umbridge} model server via \acrshort{hq}. This interface, the load balancer, is shown at the bottom of Figure \ref{fig:umbridge-setup}. It was written in C++, and dynamically registers and manages a pool of model servers distributed across a server cluster environment. We offer SLURM and \acrshort{hq} as two scheduling backends for the load balancer. This was conceptualised to function similarly to the Kubernetes load balancer, yet is specifically tailored for on-premise \acrshort{hpc} systems. 

In both backends, the load balancer acts as an intermediate abstraction layer that facilitates the deployment of concurrent model servers onto HPC compute nodes in the presence of a parallel client utilising the \acrshort{umbridge} interface. Functionally, it operates like a proxy, dynamically handling incoming client requests in a first-come, first-served manner. When requests arrive, the load balancer adaptively spawns model server instances on the HPC system by submitting jobs to one of the available backends—either SLURM or \acrshort{hq}—and subsequently routes evaluation requests to these instances. For the SLURM backend, this means directly submitting a SLURM job to a compute node with \texttt{sbatch}. For the \acrshort{hq} backend, the load balancer utilises \acrshort{hq} for the resource allocation and server task distribution instead. In a sense, \acrshort{hq} is a ``plugin'' scheduler that works on top of SLURM; it manages its own queue where the jobs are submitted and eventually distributed to the allocated nodes. The resources requested by \acrshort{hq} does not necessarily need to be a single large SLURM allocation, instead multiple shorter allocations may be requested by the interface and filled with evaluations as needed. We emphasise none of the backends interfere nor modify the behaviour of the native scheduler.

%\todo{LS: Figure 2: UQ can be on laptop, login or compute node. Load balancer and HQ MUST be on cluster though. Also, somehow indicate that we have multiple model instances. Maybe by adding multiplicity?}

In a shared \acrshort{hpc} system, spamming the scheduler queue with many jobs is discouraged because it reduces the scheduling efficiency and bloats the queue. Unfortunately, non-intrusive \acrshort{uq} methods often involve sampling, which requires submitting many small to medium jobs in a small time span\cite{Farcas2022AGF}. Launching individual SLURM jobs incurs significant overhead, making this method of submitting jobs less efficient for certain workloads. This happens, for example, when simulations exhibit irregular or unpredictable runtimes. In a SLURM script, users typically set the time limit to the maximum expected runtime, but this can introduce scheduling inefficiencies due to grossly overstating the required time limit. When a batch consists of a large number of jobs, only a few may be computationally expensive, while the majority run much more quickly. This mismatch results in underutilised resources, as the scheduler allocates resources for the full time limit, even for jobs that complete well before the maximum runtime is reached. \acrshort{hq} is able to bypass this problem by specifying a job time request in addition to a job time limit. The time request acts as a guide to the scheduler on how long each job is expected to run, whereas the time limit does not impact the scheduling, it only prevents the job from running indefinitely. 

\begin{table}[h]
\begin{threeparttable}
\caption{Main feature comparison for the three configurations of load balancers supported in the \acrshort{umbridge} framework and the standalone SLURM setup.}
    \centering
    \begin{tabularx}{0.49\textwidth}{p{1.7cm}cccc}
    \toprule
    & \multicolumn{1}{c}{UM-Bridge} & \multicolumn{1}{c}{UM-Bridge} & \multicolumn{1}{c}{UM-Bridge} &
    \multicolumn{1}{c}{SLURM} \\ 
    & \multicolumn{1}{c}{Kubernetes} & \multicolumn{1}{c}{HQ} & \multicolumn{1}{c}{SLURM} &
    \multicolumn{1}{c}{only} \\ \midrule
        Containerisation & Required & Optional & Optional & Optional \\
        Multi-node & \checkmark & Experimental & \checkmark & \checkmark \\
        Concurrent Jobs & \checkmark & \checkmark & \checkmark & \checkmark \\
        Dependent Tasks & Experimental & \checkmark* & \checkmark & \checkmark \\
        Flexible Job Time & $\times$ & \checkmark & $\times$ & $\times$ \\
        Scheduler & HA Proxy & HQ & SLURM & SLURM \\
        \bottomrule
    \end{tabularx}
    \label{tab:loadbalance_comparison}
    \begin{tablenotes}\footnotesize
    \item [*] Only available through their Python API.
    \end{tablenotes}
\end{threeparttable}
\end{table}

Table \ref{tab:loadbalance_comparison} outlines the key features and capabilities of \acrshort{umbridge} with Kubernetes, \acrshort{hq}, and SLURM configurations, compared against standalone SLURM. The comparison highlights containerisation support, multi-node support, the ability to run concurrent jobs, and support for dependent tasks. Note that flexible job times are supported only by the \acrshort{hq}-based implementation. The scheduler column identifies the primary tool used in each configuration. 

Most \acrshort{hpc} codes use \acrshort{mpi} to enable inter-node communication, allowing applications to scale across hundreds or thousands of compute nodes. For \acrshort{uq} workflows involving large-scale simulations, schedulers must provide robust multi-node support. The main challenge for the \acrshort{uq} setting is allowing another ``outer'' layer of parallelisation around the \acrshort{mpi} parallel forward model evaluations.

Certain \acrshort{uq} workflows, particularly those with parallelisable task graphs, significantly benefit from explicitly defining task dependencies within the algorithm when submitting batch jobs to a scheduler. For instance, \acrshort{uq} workflows that involve Markov Chain Monte Carlo methods require a well-defined dependency structure to manage sequential evaluations, as each step in the chain depends on the results of the previous iteration. Table \ref{tab:loadbalance_comparison} shows which load balancing approaches take task dependencies into consideration while scheduling.

\subsection{Configuration example}
This section presents brief Python code examples demonstrating three core features of \acrshort{umbridge}: setting up a model server, connecting a client to the server for making requests, and configuring a load balancer with \acrshort{hq} for efficient resource utilisation.

The model server in \acrshort{umbridge} handles requests from the client for model evaluations. Below is an example code snippet of setting up a simple server for a  simulation model:
% \begin{listing}[h]
\begin{minipage}{\linewidth}
 \begin{lstlisting}[language=Python,basicstyle=\footnotesize]
 import umbridge
 class ExampleModel(umbridge.Model):
   def __init__(self):
     super().__init__("modelname")
   [...]
   def __call__(self, parameters, config):
      """
      COMMANDS TO RUN SIMULATION
      """
   [...]
 model = ExampleModel()
 port = int(os.getenv("PORT", 4242))
 umbridge.serve_models([model], port)
 \end{lstlisting}
\end{minipage}
 % \caption{Code snippet for a setting up an UM-Bridge model server.}
 % \label{Model server code}
 % \end{listing}
This code sets up a model and serves it on port $4242$. The server listens for incoming evaluation requests.

A client can connect to the model server via \acrshort{http} and send evaluation requests. Here’s an example of a client making requests:
%\begin{listing}[h]
 \begin{lstlisting}[basicstyle=\footnotesize,language=Python]
 import argparse
 parser = argparse.ArgumentParser()
 parser.add_argument('url', metavar='url', type=str)
 args = parser.parse_args()
 model = umbridge.HTTPModel(args.url, "modelname")
 """
 CODE FOR UQ ALGORITHM
 """
 [...]
 model(input_param, config)
 \end{lstlisting}
% \caption{Code snippet for a setting up an UM-Bridge client.}
% \label{Client code}
% \end{listing}

Each load balancer job begins by executing a bash script that initialises a model server. This script generates a random port for the model server to listen on and writes the server’s network address and port information to a text file. The load balancer subsequently reads this file to register with the model server. Once registered, our new load balancer performs periodic health checks to monitor the server and distribute client requests according to the configured load balancing strategy. Below is an example of configuring \acrshort{hq} for load balancing:

% \begin{listing}[h]
 \begin{lstlisting}[basicstyle=\footnotesize,language=bash]
 #! /bin/bash
 hq alloc add slurm --time-limit=00:10:00 \
                    --backlog 1 \
                    --worker-per-alloc 1 \
                    --max-worker-count 1 \
                    -- -p shared --mem=4G 
 \end{lstlisting}
 %\caption{Bash script to launch a HQ server.}
 %\label{HQ server code}
% \end{listing}
This configuration specifies a SLURM-based allocation for a single worker, limiting the runtime to 10 minutes with 4 GB of memory. The \texttt{--backlog} option limits the number of pending requests to one, while \texttt{--worker-per-alloc} and \texttt{--max-worker-count} ensure each allocation contains a single worker. 

%%% HQ task dependency only on python api
%% and ref the table more

\section{Gyrokinetic Plasma Applications}

In this section we briefly introduce the GS2 plasma turbulence simulation and a \acrshort{gp} surrogate model thereof. Running a GS2 simulation is a fairly typical \acrshort{hpc} usage, involving large-scale, computationally intensive simulations. The resultant tasks demand significant resources but can still exhibit embarrassingly parallel characteristics in certain workloads. In contrast, predictions using the \acrshort{gp} surrogate model are computationally inexpensive, but when executed na\"ively, overheads can quickly dominate the total runtime.
These examples are intended to serve as demonstrative use cases. Our proposed method is suitable for handling problems with similar characteristics across a variety of domains. In many \acrshort{uq} workflows, both evaluations of the \acrshort{gp} surrogate and the full simulation are needed in dependence of previous evaluations, see e.g. \cite{Farcas2022AGF, lykkegaard, SCpaper}.

\subsection{GS2}

GS2\cite{GS2} is an implicit gyrokinetic plasma simulator developed to study low frequency turbulence in magnetised plasma by solving the Vlasov-Maxwell system of equations. The equation has the form 
\begin{equation}
\frac{\partial F_a}{\partial t} + \frac{\partial \Vec{X}}{\partial t} \cdot \nabla F_a + \frac{\partial v_\parallel}{\partial t} \frac{F_a}{\partial v_\parallel} = 0,
\end{equation}
where $F_a = F_a(\vec{x}, v_\parallel, \mu)$ is the gyrocenter distribution function for
species $a$ represented in full 5D phase space, $\vec{x}$ is the three-dimensional vector of coordinates of the guiding center of a particle, $v_\parallel$ is the velocity coordinate along the magnetic field line, and $\mu = v^2_\perp / 2B$ is the magnetic moment (a conserved quantity), where $v_\perp$ is the velocity in the perpendicular direction and $B$ is the magnetic field strength. It is mainly used for calculating key properties of a turbulent plasma to improve understanding of microinstabilities in plasmas produced in a laboratory.

There are two ways to initialise GS2, either as an initial value solver where the simulation ends the moment an unstable mode is found or as an eigenvalue solver where it finds all the eigenvalue pairs for a given parameter set; all of the experiments presented are linear and ran in the initial value solver mode. Due to this, the runtime depends heavily on the input parameters as some are more unstable than others—it varies between the order of minutes to hours using a compute node with two AMD EPYC 7702 64-cores processor in some cases. Furthermore, the simulation runtime is not easily predicted for a given set of inputs. The associated computational challenges motivated the development of a reliable reduced order model using a \acrshort{gp}\cite{10.1063/5.0174478, Rodriguez-Fernandez_2024, shinya_2024}. Our GS2 benchmark closely follows the simulation as it was ran in the cited work, where the authors varied seven input parameters to find unstable modes known as micro-tearing modes. The description of the inputs is shown on Table \ref{GS2 inputs table}. We sample from the same set of input parameters using the Latin Hypercube sampler. However, our benchmark converges to a different unstable mode known as a kinetic ballooning mode. These modes are more relevant to Tokamak transport modelling and can be run at lower resolution leading to a decreased computational cost.% The time to solution for this benchmark, varies from a few minutes to hours using 8 CPU cores depending on proximity to the unstable region.

\begin{table}[h]
\centering
\caption{The input parameters varied in the GS2 simulations.}
\begin{tabular}{lcc}
\toprule
\multicolumn{1}{l}{Input name} & \multicolumn{1}{l}{Minimum} & \multicolumn{1}{l}{Maximum} \\ \midrule
Safety factor & 2  & 9      \\
Magnetic shear    & 0 & 5  \\
Electron density gradient   & 0  & 10    \\
Electron temperature gradient   & 0.5  & 6    \\
Ratio of plasma pressure to magnetic pressure     & 0  & 0.3  \\
Electron–ion collision frequency                & 0  & 0.1   \\
Bi-normal mode wavelength           & 0 & 1  \\ \bottomrule
\end{tabular}
\label{GS2 inputs table}
\end{table}

% Find somewhere to put this
%The application was designed to be a parallel code that also scales well across multiple HPC systems (insert figure form documentation maybe?). Distributed parallelism (MPI) is preferred for running the simulation though shared memory parallelism is also supported.

\subsection{Gaussian Process Surrogate} 
A \acrshort{gp} is typically used as a surrogate model for complex and lengthy numerical simulations. This approach has gained traction recently over other conventional black-box methods (e.g. neural networks) due to its interpretability and rigorous treatment of uncertainties coming from Bayesian statistics.

Formally, a \acrshort{gp} \cite{rasmussen2006gaussian, article} is defined as a stochastic process (a collection of random variables) such that any finite number of which have a joint Gaussian distribution, i.e. the function value $f(\mathbf{x})$ for each input vector $\mathbf{x}$ is a random variable satisfying the above property. Alternatively, identical results can be reached by considering in terms of a Bayesian linear regression setting $f(\mathbf{x}) = \phi(\mathbf{x})^\top \mathbf{w}$ for some basis function $\phi(\mathbf{x})$ with a Gaussian prior placed over its weights $\mathbf{w}$.

%\todo[inline]{Mikkel: Not exactly. We kind of have to write $f(\mathbf{x}) = \phi(\mathbf{x})^\top \mathbf{w}$ with some particular conditions of $\phi$ for that equivalence to hold, namely that $k(x, x') = \phi(x)^\top\Sigma_0\phi(x')$. Otherwise GP regression would be far too easy :)}

\begin{figure}[h]
    \centering
    \includegraphics[width=0.4\textwidth]{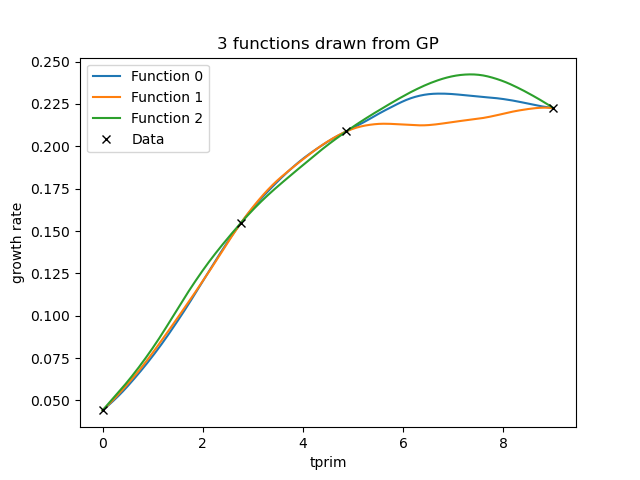}\label{gp-function}\\
    \includegraphics[width=0.4\textwidth]{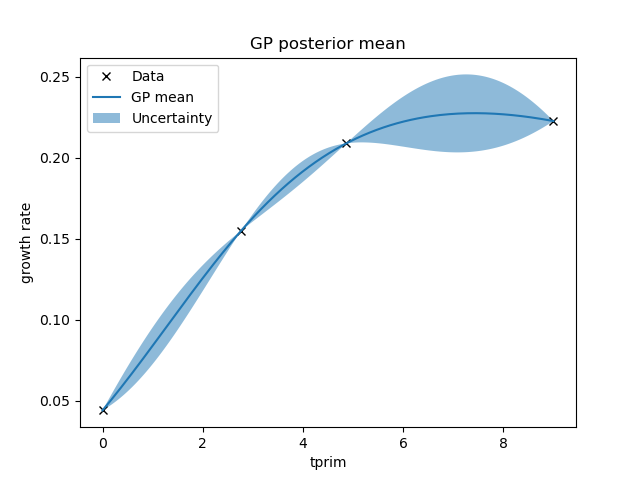}\label{gp-mean}
    \caption{Top: Three functions drawn from a \acrshort{gp} posterior distribution where $\times$ are 4 training data points. Bottom: Mean and uncertainty obtained from the trained \acrshort{gp}. Again, $\times$ represents 4 training data points, and the shaded blue region corresponds to the 95\% confidence interval.}%\todo[inline]{Mikkel: Just a semantic point, I mildly prefer "posterior" GP over "trained" GP just to emphasise the Bayesian connection. "Mean can be understood as the average of the functions drawn." is unnecessary, since we explain it in the text.}}
    \label{fig:gp}
\end{figure}

In both cases, the \acrshort{gp} is written as 
\begin{equation}
f(\mathbf{x}) \sim GP(m(\mathbf{x}), \text{cov}(\mathbf{x}, \mathbf{x'})),
\end{equation}
where $m$ is the mean function to provide a-priori structure from the dataset, and $\text{cov}(\mathbf{x}, \mathbf{x'})$ is the covariance function (also known as a kernel) that measures the covariance of pairs of points in the input space and map them into a semi-positive definite matrix. These two functions completely specifies a \acrshort{gp}, though the mean function is often set to zero or other constant for simplicity. On the other hand, the choice of kernel strongly impacts the quality of the prediction. Once conditioned on observed data, the mean of the posterior probability distribution $\bar{f}(\mathbf{x_*})$ gives the output prediction at input $\mathbf{x_*}$, whereas the standard deviation sets the uncertainty away from the mean. Figure \ref{fig:gp} shows an example figure of a GP conditioned on artificially generated data points. Assuming the observations $y$ were corrupted by independent and identically distributed additive Gaussian noise with standard deviation $\sigma_n$, the mean and variance of the posterior distribution $\mathbb{V}$ are given by
\begin{align}
\begin{split}
\bar{f}(\mathbf{x_*}) ={}& k(\mathbf{x}, \mathbf{x_*})^\top (k(\mathbf{x}, \mathbf{x}) + \sigma^2_n I)^{-1} y(\mathbf{x}), 
\end{split}\\
\begin{split}
\mathbb{V}[f(\mathbf{x_*})] ={}& k(\mathbf{x_*}, \mathbf{x_*}) \\
& - k(\mathbf{x}, \mathbf{x_*})^\top (k(\mathbf{x}, \mathbf{x}) + \sigma^2_n I)^{-1} k(\mathbf{x}, \mathbf{x_*}),
\end{split}
\end{align}
%\todo[inline]{Mikkel: Actually $y(\mathbf{x_*})$ is unknown a-priori, so the last factor in the expression for the mean is just $\mathbf y$ which is the \textit{training} outcomes.}
where $I$ is the identity matrix and $k(\mathbf{x}, \mathbf{x_*})$ is the kernel measuring the correlation between data points $\mathbf{x}$ and $\mathbf{x_*}$. In addition to conditioning the \acrshort{gp} model to the experimental data, the dominant computations here are matrix operations on $k$, notably matrix inversion. 

In a similar spirit to the GS2 benchmark, the \acrshort{gp} model we use for benchmarking is pre-trained and derived from \cite{10.1063/5.0174478}, where it accepts the same 7 input parameters and outputs 2 scalar quantities—namely, mode growth rate and mode frequency.

\subsection{Computing the Quantity of Interest}\label{sec:qoi}
The end goal in our use case is the computation of multiple integrals for a set of high-dimensional input parameters. The integrals of interest in our application have the form:
\begin{multline}
    Q_{\mathrm{ql},s} = Q_0 \Lambda^{\alpha-1}\biggl(\frac{1}{\rho_* c_s}\biggr) \\
         \int\mathrm{d}k_y\;\frac{1}{\theta_{0,\mathrm{max}}}\int_0^{\theta_{0,\mathrm{max}}} \mathrm{d}\theta_0 \frac{Q_{\mathrm{l},s}(k_y,\theta_0)}{Q_{\mathrm{l}}(k_y,\theta_0)} \hat{\Lambda}(k_y, \theta_0),
\label{eq:quantity_of_interest}
\end{multline}
 which calculates a quantity of interest typically found in nuclear fusion research\cite{Kumar_2021}. This is an example of a quasi-linear saturation rule where non-linear fluxes are approximated using the results of linear simulations. More details on this integral are presented in \cite{integral}, see equation (3.6).

\vspace{-0.25cm}

 We can approximate (\ref{eq:quantity_of_interest}) with quadrature rules or Monte Carlo methods. However, computing (\ref{eq:quantity_of_interest}) can become infeasible since the accuracy of the numerical solution depends on the number of evaluated points. By approximating the integrand with a \acrshort{gp} surrogate, we can control the cost by balancing the prediction uncertainty of the \acrshort{gp} and the precision of the numerics. It is also common to couple the \acrshort{gp} surrogate model with an acquisition function to make informative guesses in exploring the parameter space to make adaptive update to the \acrshort{gp}. Similar ideas are known as Bayesian quadrature, and explored in \cite{pmlr-v108-fisher20a, Hennig_Osborne_Kersting_2022}.

The purpose of the integral evaluations in our experiments is to demonstrate how quickly the forward map or \acrshort{gp} surrogate can compute these values, as a proxy for the efficiency of our scheduling approach. Since this is not a fusion-focused study, we do not present the final results of the integral computations. Instead, the evaluations in the results section illustrate the potential speedups that can be obtained by improving scheduling efficiency of the necessary evaluations of GS2 and the \acrshort{gp} surrogate.

%%% say software versions
\section{Performance Testing} \label{performance testing}
All benchmarks were run on the Hamilton8 supercomputer hosted by Durham University. The code and experimental data are publicly available on \href{https://github.com/chun9l/UMBridge_Loadbalancer}{GitHub\footnote{\url{https://github.com/chun9l/UMBridge_Loadbalancer}}}. The system operates on Rocky Linux 8, and provides a total of 15,616 CPU cores, 36TB RAM and 1.9PB disk space, spread over 122 compute nodes (120 standard nodes plus 2 high memory nodes), which are connected via Infiniband HDR 200GB/s high performance interconnect, with a 2.6:1 fat tree topology formed of non-blocking islands of up to 26 nodes. We run our experiments solely on the standard nodes. Each standard nodes consists of 128 CPU cores (2x AMD EPYC 7702), 256GB RAM (246GB available to users) and 400GB local SSD storage. Hyperthreading is disabled across all nodes to minimise resource contention between threads for the consistency and reliability of results. 

The Hamilton8 supercomputer is active with approximately 60 online users and 700 running jobs at the time of testing. We did not request exclusive access or special priority queues, ensuring that our jobs were treated the same as any other submission. The experiments were spread out and performed over several days to ensure subsequent submissions have the same priority. This is because the SLURM on our system deprioritises a user's submissions once they have reached a certain number of submissions. At its core, the \acrshort{hq} implementation relies on the native SLURM installation in the system, so the deprioritisation of jobs applies there as well.

%%% mention we do not use pin cpu, and use default mpi binding and mapping policy?
%%% In a similar spirit to the container scaling test in Section \ref{}, we leave the OpenMPI binding and mapping policy as default instead of manually pinning the CPU cores. 
%%% Maybe go into appendix: Using Singularity 4.1.5-1.3l8, latest Ubuntu 20.04 from Dockerhub, Python3.9.19 managed by conda 24.1.2, numpy version 1.26.4, umbridge 1.2.4. GCC 11.2.0 and Openmpi 4.1.1 on host. HQ v0.20.0 and GS2 8.1.2 at commit 89d63c2.

Additionally, we wish to highlight several issues encountered on the Hamilton8 
system, along with their corresponding workarounds.
Currently, \acrshort{umbridge} relies on a text file to communicate the IP address and port number of the model running on the compute node to the load balancer. One issue observed was that the text file, although written, was not visible to the load balancer. This was found to be due to the filesystem not updating in a timely manner. To address this, we manually integrated the \texttt{sync} command into the load balancer's source code, ensuring that the filesystem is refreshed before the file is accessed. 
%Another issue occurred where the text file was created, but it appeared empty when the load balancer attempted to read it. This was resolved by modifying the load balancer's source code to verify that the file was not empty before attempting to retrieve the connection information. 
These issues are likely specific to our testbed, possibly related to inadequate filesystem caching or the I/O-intensive nature of the tasks conducted during the experiments. Our collaborators did not encounter these problems on the Helix 
supercomputer at Heidelberg University, where similar experiments were conducted.

%%% This decision was made to eliminate any potential variable performance effects that SMT may introduce as a result of possible resource contention. This is a typical configurations in HPC clusters, where it is standard to run as many processes as there are cores on a machine in order to maximise resource utilisation. %ref?
%This was copied from Darius dissertation EDIT: Looked up some literature, couldn;t find a conlusive answer to this. It benefits some applications but also worsen some. I think just mention Hyperthreading is disabled.

\subsection{Metrics} 
The simplest way to compare the performance between schedulers is by looking at the total execution time of the respective benchmark. While this clearly indicates which scheduler yields better performance, it is specific to the finished benchmark only, i.e. non-transferable, and does not directly expose the scheduling overhead which is the focus of this paper. 

There are many ways to express this quantity, most of them involve comparing the CPU utilisation time and a measured runtime. For simplicity, the total runtime of a job (makespan) is assumed to be separable into two mutually exclusive additive parts: scheduling overhead, and CPU time. To clarify, this CPU time is defined for the job submitted to the scheduler rather than only at the application, this means the timer begins when the job starts. Additionally, we deliberately include the queueing time into the scheduling overhead, but one can argue that the inclusion of queuing time is inappropriate because it is dependent on the system utilisation. Indeed, resource allocation is easier in an empty cluster than one which is running at full capacity. The scheduler's responsibility is to allocate resources to submitted jobs regardless of the system utilisation, hence we include queuing time as a part of the overhead.

In \cite{993206} a list of useful comparison metrics is presented, one of which is the \acrfull{slr}. We adopt this metric to compare the efficiency of different schedulers. The \acrshort{slr} is defined by
$$
SLR = \frac{makespan}{\sum_i C_i},
$$
where $makespan$ is the length of the output schedule, and $C_i$ is the compute time of the $i^{th}$ task. \acrshort{slr} represents the actual elapsed time utilisation as a multiplier of program CPU time utilisation. For instance, a factor of 1.0 implies perfect utilisation—i.e. the lower bound with zero scheduler overhead, whereas a factor of 3.0 implies that the total elapsed time was 3x longer than expected—i.e. the scheduler took 2x longer than the total program execution time in its execution overhead. 

\subsection{Benchmarks}
As a reference, we use Python scripts to pseudo-load-balance the job submissions, independent of \acrshort{umbridge}, because this is the predominant method among users of the fusion simulator GS2 who submit a batch of 1000 jobs at a time with this method. We chose not to use the SLURM backend in \acrshort{umbridge} to compare directly against the \acrshort{hq} backend because it was designed as a simpler alternative. Since it submits individual SLURM jobs without altering the core scheduling mechanism, there is no performance gains over our baseline approach and was not included in our performance evaluation. Nevertheless, we provide some benchmark results from the SLURM backend in Appendix \ref{appendix}.

\begin{table}[h]
\centering
\caption{The resources requested by each benchmark.}
\begin{tabularx}{0.49\textwidth}{p{3cm}cccc}
\toprule
 & \multicolumn{1}{c}{eigen-100} & \multicolumn{1}{c}{eigen-5000} & \multicolumn{1}{c}{gs2} & \multicolumn{1}{c}{GP} \\ \midrule
SLURM Allocation Time (mins) & 1  & 5  & 240 & 1   \\
HQ Allocation Time (mins)    & 10 & 60 & 36000 & 10 \\
HQ Job Time Request (mins)   & 1  & 5  & 15  & 1 \\
HQ Job Time Limit (mins)     & 5  & 10 & 240 & 5 \\
SLURM/HQ CPUs                 & 1  & 1  & 8  & 1  \\
SLURM/HQ RAM (GB)            & 4  & 4  & 32  &  4 \\ 
Expected time to solution (mins) & $0.01$ & 2 & [1,180] & $0.1$ \\
\bottomrule
\end{tabularx}
\label{resources table}
\end{table}

In addition to the GS2 and \acrshort{gp} evaluations, we present a simpler problem that computes the eigenvalues and the corresponding right eigen-vectors of a randomly generated square matrix. The Numpy \cite{harris2020array} function \texttt{numpy.linalg.eig} is employed to solve the eigenproblem, this function calls the solver \texttt{\_geev} from LAPACK \cite{laug} behind the scenes. Note that this benchmark is memory bound since the matrices are not sparse. However, these benchmarks are still comparatively cheap and easy to compute. We refer to the problem of solving for a size $100$ and $5000$ system as  \texttt{eigen-100} and  \texttt{eigen-5000} respectively. 

For each example application, we perform two sets of experiments consisting of 100 evaluations on both schedulers, where either two or ten jobs were allowed in the queue, i.e. we maintain a fixed number of jobs in the queue throughout the course of the experiment. This intends to mimic the behaviour of a user submitting jobs one after the other, up to a predefined threshold. For the purpose of clarity, we refer a benchmark as the set of 100 evaluations performed on an example application. The series of evaluation in each benchmark were generated with the same random seed for repeatability across different runs. For example, matrices in the \texttt{eigen-100} benchmark are the same for all 100 evaluations, and the input parameters for GS2 are sampled from a seeded Latin hypercube sampler. Hence, runtime variations from any repeated benchmark do not originate from the benchmark problem. These fluctuations are related to the hardware itself as well as the load of the cluster at the time of the experiment.

The resources requested by an application are constant throughout the benchmarks, doing so enforces fairness of the experiments. As the time request works differently in SLURM and \acrshort{hq}, we set the time limit in SLURM to be the expected maximum runtime. There are 3 time settings in \acrshort{hq} that we specify: allocation time limit, job time limit, and job time request, where the latter two were mentioned in Section \ref{load balancer}. We pass the total expected runtime (for all 100 runs) as the allocation time, then the job times are set so that the time request corresponds to the minimum expected runtime for each iteration, and the time limit to the maximum expected runtime. We add a buffer of additional time to these time requests to allow for runtime variation. Table \ref{resources table} shows the actual resources request submitted to the respective scheduler for each benchmark.

\section{Scheduler Comparison}
%%% Add some comments regarding spread of results

%\todo{refer to 2 node and 10 node instead?}

%\begin{figure}[]
%    \centering
%    \includegraphics[width=0.4\textwidth]{scheduler_plots_gp/slurm_hq/makespan_10.pdf}
%    \includegraphics[width=0.4\textwidth]{scheduler_plots_gp/slurm_hq/CPUTime_10.pdf}
%    \includegraphics[width=0.4\textwidth]{scheduler_plots_gp/slurm_hq/Scheduler Overhead_10.pdf}
%    \caption{Boxplots showing experimental results with 10 jobs filling the queue.}
%    \label{10 jobs}
%\end{figure}

\begin{figure*}[]
    \centering
    \includegraphics[width=0.4\textwidth]{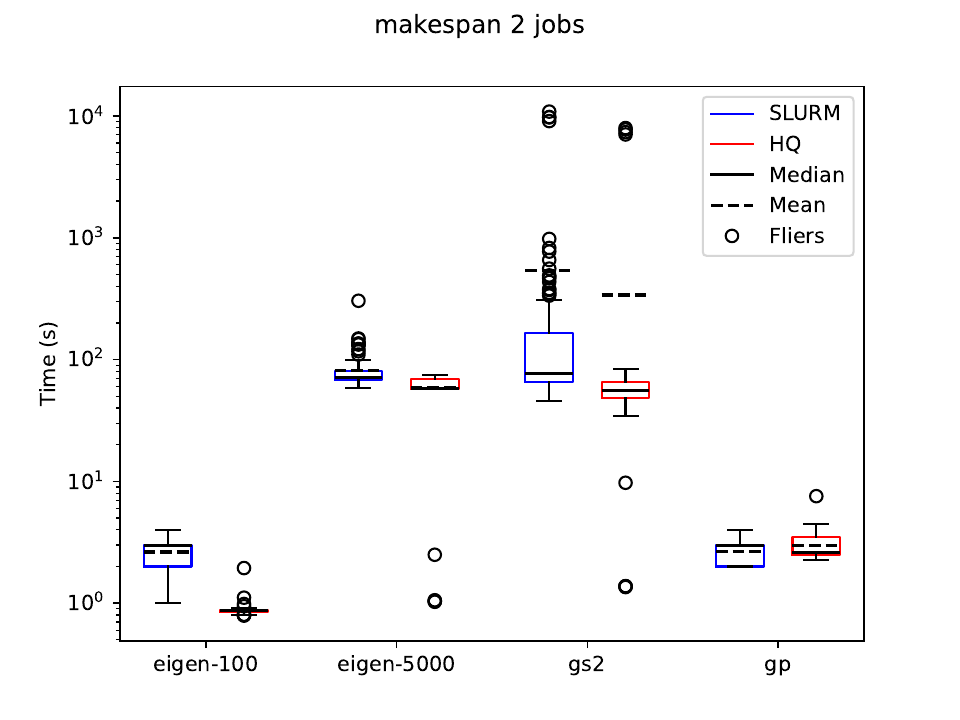} \includegraphics[width=0.4\textwidth]{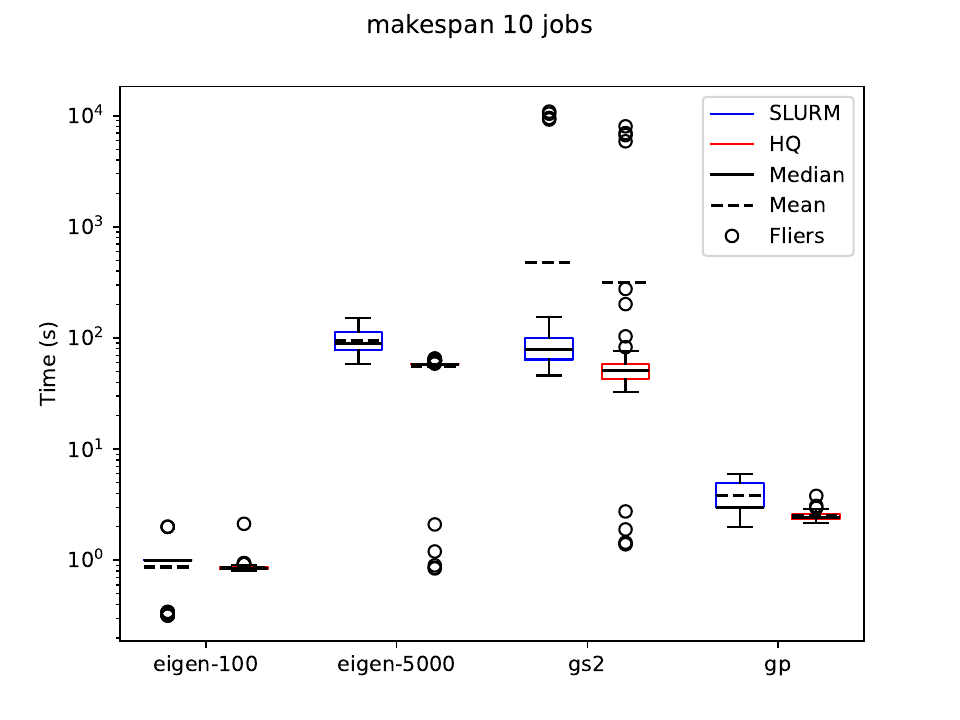}
    \includegraphics[width=0.4\textwidth]{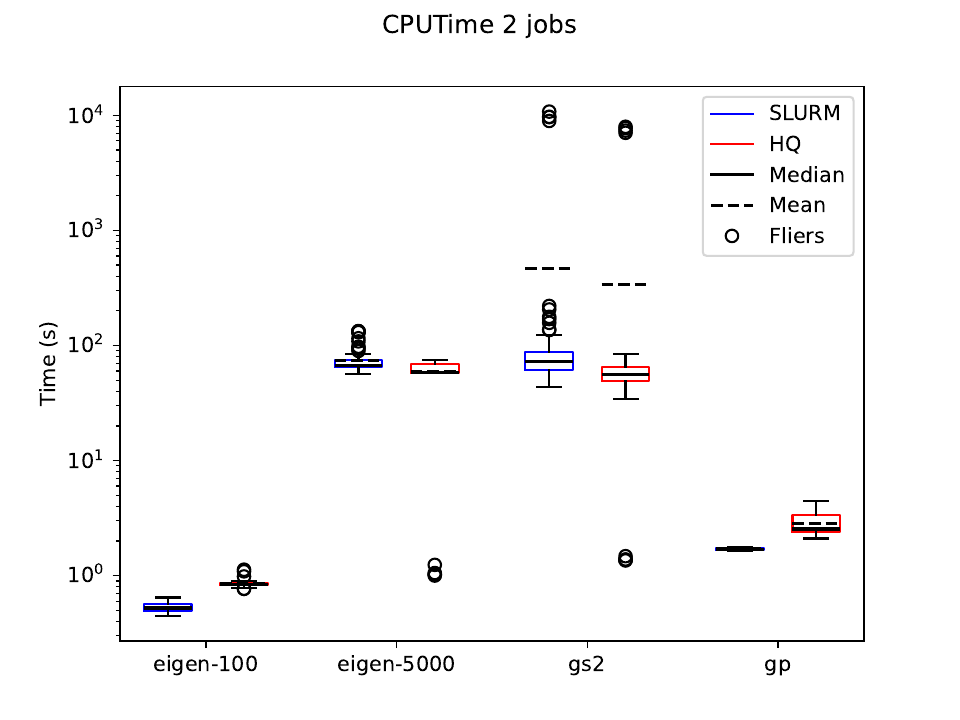}\includegraphics[width=0.4\textwidth]{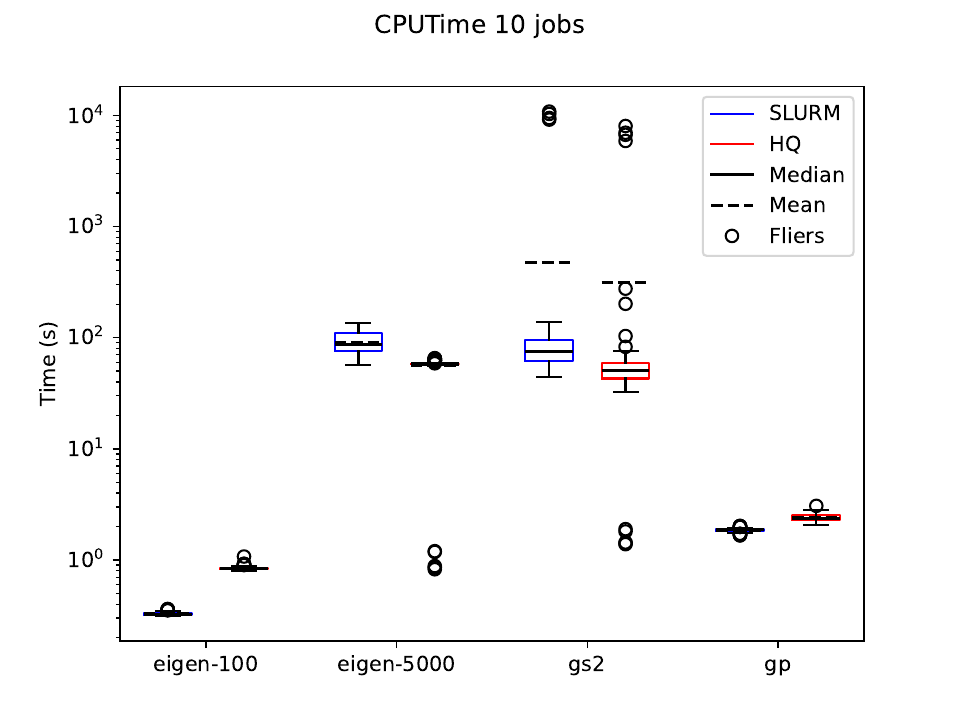}\\
    \includegraphics[width=0.4\textwidth]{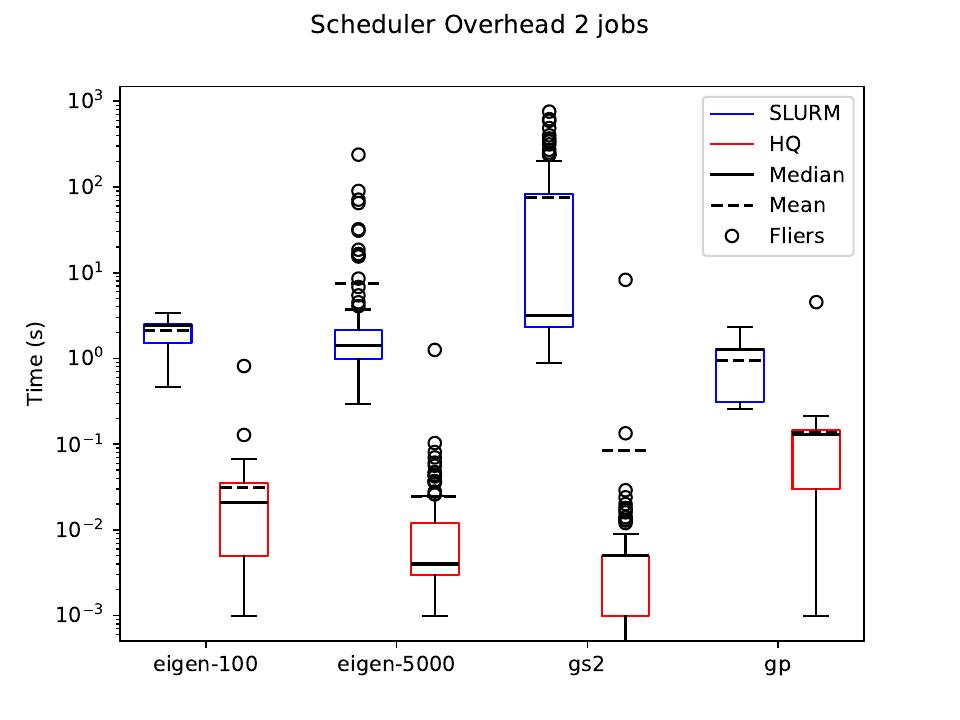}\includegraphics[width=0.4\textwidth]{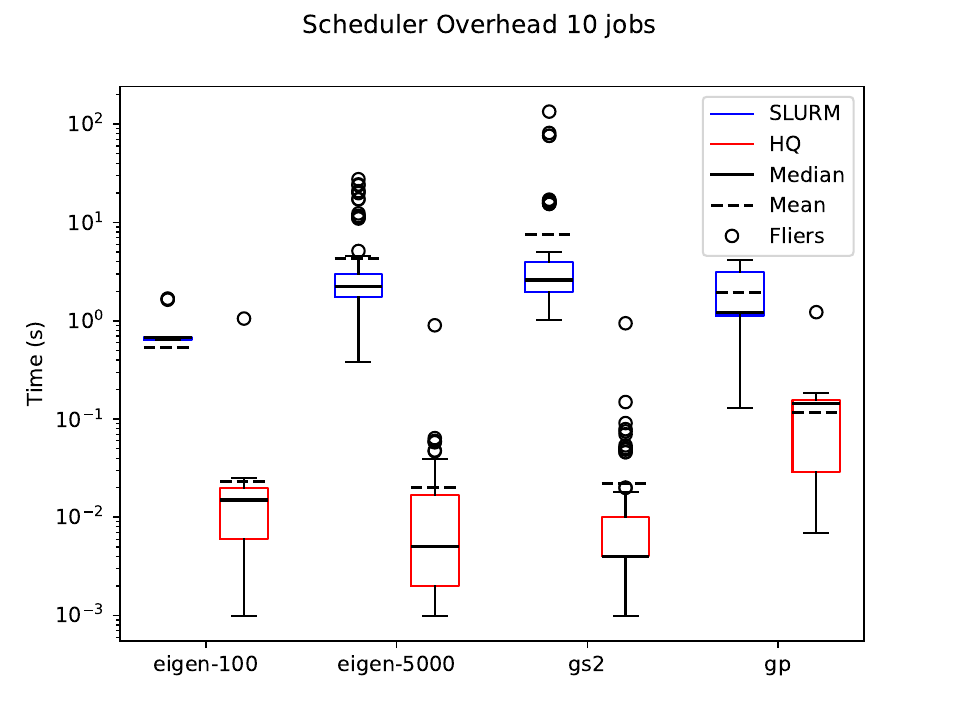}\\
    \caption{Boxplots showing experimental results with 2 jobs (left column) and 10 jobs (right column) filling the queue. For each application (listed on the x-axis), the left (blue) boxes represent data collected from SLURM and the right (red) boxes represent data from \acrshort{hq}. The top row shows the makespan, the middle row the CPU time, and the bottom row shows the scheduler overhead, all measured in seconds.}
    \label{2 jobs}
\end{figure*}

\begin{figure}[h]
    \centering
    \includegraphics[width=0.4\textwidth]{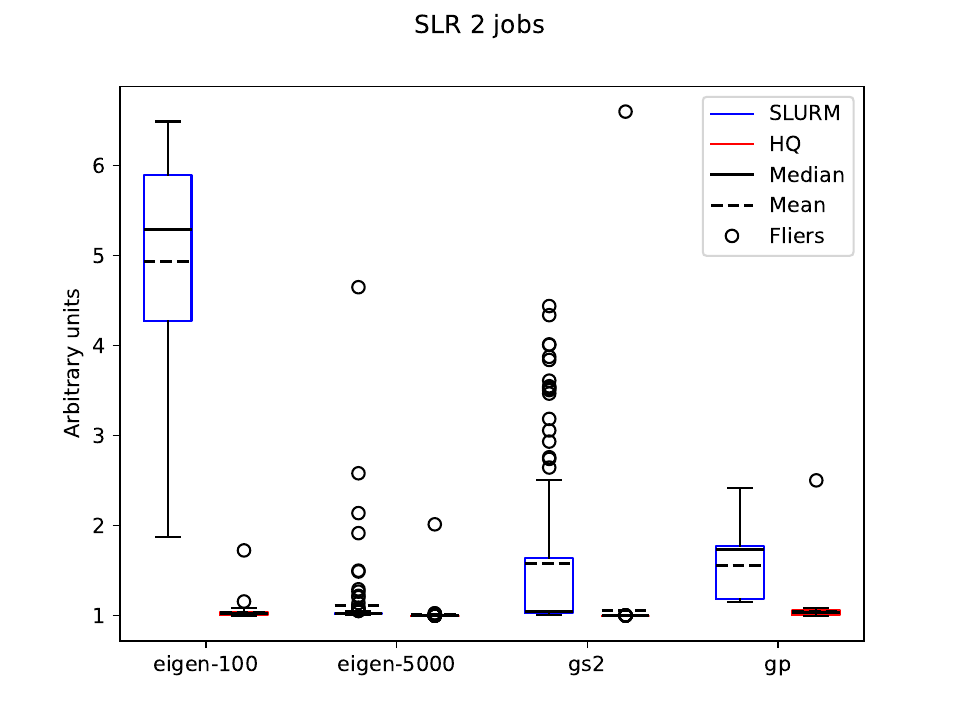}
    \includegraphics[width=0.4\textwidth]{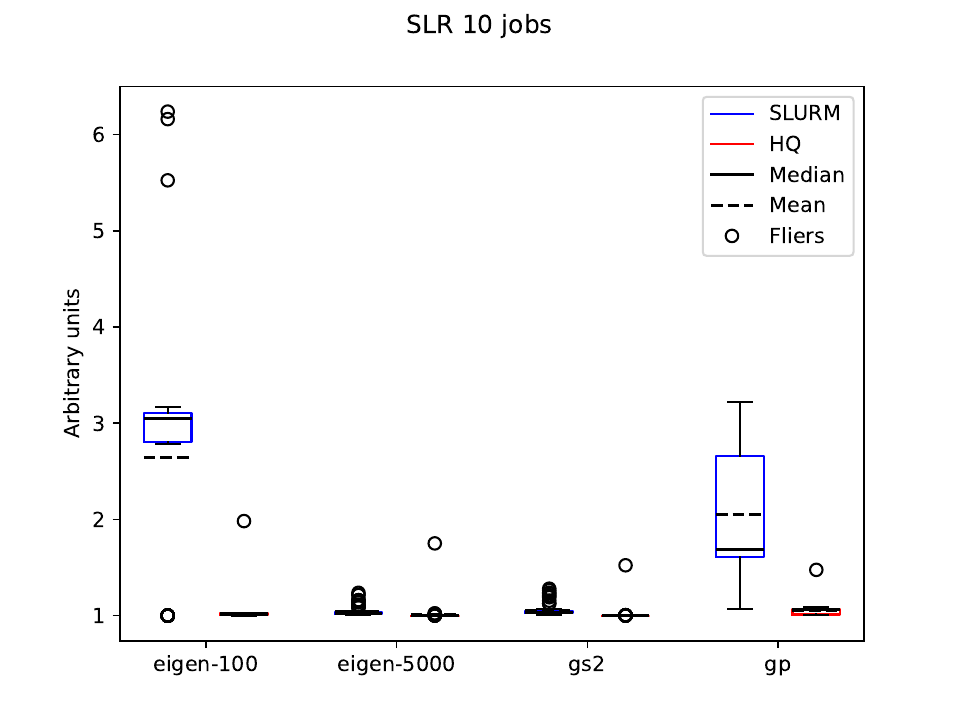}
    \caption{Boxplots showing the SLR for two jobs filling the queue (top) and 10 jobs filling the queue (bottom). For each application (listed on the x-axis), the left (blue) boxes represent data collected from SLURM and the right (red) boxes represent data from HQ.}
    \label{slrs}
\end{figure}

We present our experimental results as boxplots in Figures \ref{2 jobs} and \ref{slrs}. In each of these figures, results for the four different test cases are shown. Each application is listed on the x-axis with the left (blue) boxes representing data collected from SLURM, while the right (red) boxes represent data from \acrshort{hq}. The six plots in Figure \ref{2 jobs}  provide a detailed breakdown of the timings discussed in Section \ref{performance testing}, and Figure \ref{slrs} displays the \acrshort{slr} for the two types of experiments conducted, i.e. for the cases of two and ten jobs filling the queue respectively. 

All times were obtained from the native logs recorded by SLURM and \acrshort{hq} respectively. \acrshort{hq} logs all of its timings with an accuracy to milliseconds. Unfortunately, SLURM only records with granularity of up to seconds, with the exception of CPU time where it is accurate to microseconds. We derive the scheduler overhead by subtracting the CPU time from the makespan. Thus, extra checks were needed for logs produced by SLURM to prevent getting erroneous results such as a negative overhead. If the run is fast enough that the makespan is zero, we set it to the CPU time and assume zero scheduler overhead instead.

On average, the makespans indicate the majority of the experiments ran with \acrshort{hq} finished first—though most of them were only quicker by a relatively small margin. The most significant reduction occurs at \texttt{eigen-100} with 2 jobs, where the \acrshort{hq} approach is roughly three times quicker. Although our claim that the \acrshort{hq} version is better suited for quicker jobs, e.g. \texttt{eigen-100}, aligns with the experimental outcomes, it is outperformed by SLURM in terms of CPU time. This is due to the model initiation overhead accompanied in each \acrshort{hq} job, where the model server takes approximately 1 second (regardless of the application) to start up before the load balancer/client can connect to it. These delays are most obvious when comparing the CPU time between the schedulers in shorter runs like \texttt{eigen-100} where the time to solution is less than 1 second. In longer jobs, they are masked by the actual computation of the application. Nevertheless, the \acrshort{hq} version achieves a lower overall runtime for most test cases.

%%% possibly add another paragraph about running gp without Umbridge

In the \acrshort{hq} implementation, at least five additional jobs are consistently submitted due to the load balancer’s design. These jobs arise from the load balancer querying the model for information prior to processing the first evaluation request, with each job incurring approximately one second of server initialisation overhead. While these jobs do not contribute directly to the computation of the benchmark, they play a role in enhancing the reliability of the workflow. By verifying the readiness of the model server and ensuring both client and server expect the correct input and output dimensions, these preliminary jobs help mitigate potential runtime issues that could otherwise disrupt the execution. For longer-running applications such as GS2, these additional jobs appear as lower outliers in the boxplot. In faster applications, however, they blend into the typical runtime range and are less noticeable. In contrast, the SLURM implementation does not exhibit this behaviour, as it operates independently of \acrshort{umbridge}.

Interestingly, \acrshort{hq} runs exhibit lower CPU time than SLURM in longer-running applications such as GS2. This result is somewhat unexpected, as the CPU time should primarily reflect the computational workload of the application, which we expect to be similar across schedulers. Additionally, \acrshort{hq} incurs the server launch overhead of around one second per submission, which theoretically places it at a disadvantage. Upon closer examination, this discrepancy can be attributed to differences in how the two schedulers allocate and utilise compute resources. In the native SLURM submission, an allocation request is made for every job independently, with no guarantees about how these jobs are distributed across compute nodes. Consequently, SLURM must reinitialise the environment for each job, leading to additional overhead that is reflected in the CPU time. This contrasts with \acrshort{hq}, which receives distinct nodes in a single allocation request that persist throughout the length of the experiment, thus avoiding repetitive setup costs. SLURM’s tendency to assign multiple jobs to the same node introduces variability. When several jobs are executed on the same node, simultaneous filesystem access and resource contention potentially increase CPU time. This behaviour could also contribute to greater variation in SLURM runtimes observed in Figure \ref{2 jobs}. In contrast, \acrshort{hq}'s allocation strategy leads to a more consistent CPU time.

Because \acrshort{hq} only request one SLURM allocation, its primary overhead is waiting for SLURM to allocate the requested resources. This is the first job submitted to \acrshort{hq}, these jobs are consistently the biggest contributor to the scheduling overheads, as well as the highest valued \acrshort{slr}s on Figure \ref{slrs}. After that, the job launch overhead diminishes to the order of milliseconds. On the other hand, SLURM carries considerable overhead for launching each job after receiving the requested resources. Thus, the \texttt{eigen-100} testcase performed better in \acrshort{hq} because the reduction in scheduler overhead outweighs the gain in CPU time.

The \acrshort{slr} as a metric clearly accentuates the performance of \acrshort{hq} because the overhead in each \acrshort{hq} job is insignificant, i.e. the makespan and CPU time are approximately equal. Note that this metric only considers overhead from the scheduler which is not indicative of the overall performance. One may also argue that we reduced the scheduling overhead but injected other overheads into the CPU time (as seen in fast running benchmarks in Figure \ref{2 jobs}). However, in all but the fastest running testcases, the overall runtime decreases. In the case of the costly GS2 simulation, the mean makespan has decreased by around 38\% for both the 2 and 10 jobs setting. Only in the case of very fast running jobs there is a slight increase in runtime. This inefficiency could potentially be removed by introducing persistent servers for fast running jobs.

%We do not observe significant differences between 10 or 2 concurrent jobs. This is likely because 10 parallel jobs are still considered ``light load'' by SLURM. We need to maintain higher job counts to stress the scheduler and see more pronounced effects.
% \todo[inline]{this opens the question of why we haven't done that...}

%%% Improvements: current architecture not optimal eg io dependent, overhead from starting server, port race condition

\section{Conclusion}

%We integrated the UM-Bridge interface with HQ, and deployed it to run problems of various workloads. This framework is further tested against the na\"ive way of scheduling these jobs using SLURM installed on our local HPC system. The results demonstrate that the HQ version performs better, if not similar, compared to the SLURM approach across the example applications, especially for jobs with short runtime. This is due to SLURM incurring considerable overhead for each resource request, whereas the HQ approach minimises this by submitting one bulk request in the beginning. Again, we stress that our UM-Bridge framework is not restricted to the examples presented here, it can be applied to problems of the same nature.

In this paper, we demonstrated the newly developed \acrshort{umbridge} load balancer for classical \acrshort{hpc} systems. We tested this \acrshort{hq}-based load balancer on realistic workloads from a gyrokinetic plasma application and compared its performance with the traditional SLURM scheduler on our local \acrshort{hpc} system Hamilton8. Our results show that the \acrshort{hq}-based approach either outperforms or is comparable to SLURM for both fast-running and compute intensive tasks. This is largely due to SLURM incurring considerable scheduling overhead for each resource request, whereas \acrshort{hq} minimises this by submitting one bulk request in the beginning. The latter approach reduced the scheduling overhead by up to three orders of magnitude. 

%The current UM-Bridge HQ-based framework already has most of the core features implemented. However, it will benefit from several architectural improvements. Recall, the cost of initialising model servers per job is a bottleneck to achieve higher performance, this is avoidable by implementing a persistent server that can be reused repeatedly over the duration of the allocation. Other than that, we are working towards a more robust network-based method to relay connection metadata of the UM-Bridge model rather than writing them to a text file.
The load balancing approach introduced in this paper has all of the necessary core features allowing users to easily scale their \acrshort{uq} applications across large clusters. We emphasise that the framework is not restricted to the examples presented and can be adapted to a wide range of applications with similar characteristics, including loosely-coupled, parallel tasks. However, there are several potential areas for architectural improvement. Recall, the cost of initialising model servers per job is a bottleneck to achieve higher performance, this is avoidable by implementing a persistent server that can be reused repeatedly over the duration of the allocation. To address issues related to filesystem dependencies, we are working towards a more robust network-based method to relay connection metadata of the \acrshort{umbridge} model rather than writing it to a text file. 
%Once the proposed improvements for UM-Bridge has been implemented, we will revisit these benchmarks to investigate the changes in performance, as well as increasing the concurrent job limit to stress the scheduler. % see whether the test go through 

The main area for future exploration is fully testing the capability of the framework to handle more complex workflows, where tasks have interdependencies or dynamic scheduling requirements. In particular, we are interested in deploying this framework to compute the integral \eqref{eq:quantity_of_interest} with an adaptive \acrshort{gp} model. The computational workload in this example varies vastly as it involves evaluations of a costly simulation and significantly cheaper predictions from the \acrshort{gp} surrogate. Furthermore, the adaptivity of the \acrshort{gp} model brings in additional complexity to the workload by introducing loosely dependent tasks, which contrasts with the embarrassingly parallel examples presented in this paper. Successful realisation of this method will streamline the process of scheduling complicated task graphs, alongside delegating costly simulation to the surrogate at points with low uncertainty.

Lastly, we acknowledge that the example applications presented may not be representative of all \acrshort{uq} workloads, and may even be cheap compared to more costly simulations. We chose the GS2 and \acrshort{gp} combination since we were primarily interested in improving the scheduling of tasks with reasonably varying runtimes. As this framework matures, we would be interested to incorporate a more diverse application mixture in a follow-up work.

\section*{Acknowledgments}
We thank Tuoxing Liu and Lev Gromov for their significant contributions to the UM-Bridge \acrshort{hpc} load balancer during their software practicals at Heidelberg University and Darius Plesan Tohoc for his contributions while working on his BSc thesis at Durham University. We thank Jakub Ber\'anek and Ada B\"ohm for their very helpful discussions on the usage of HyperQueue. 

The authors acknowledge support by the state of Baden-Württemberg through bwHPC and the German Research Foundation (DFG) through grant INST 35/1597-1 FUGG.

%\todo[inline]{LS: I'll have to report the paper to bwHPC (once published) for their stats}

\bibliographystyle{IEEEtran}
\bibliography{refs.bib}

\clearpage

\appendices 
\section{Supplementary Figures} \label{appendix}
We provide additional figures here to complement the main text. 

\begin{figure}[h]
    \centering
    \includegraphics[width=0.4\textwidth]{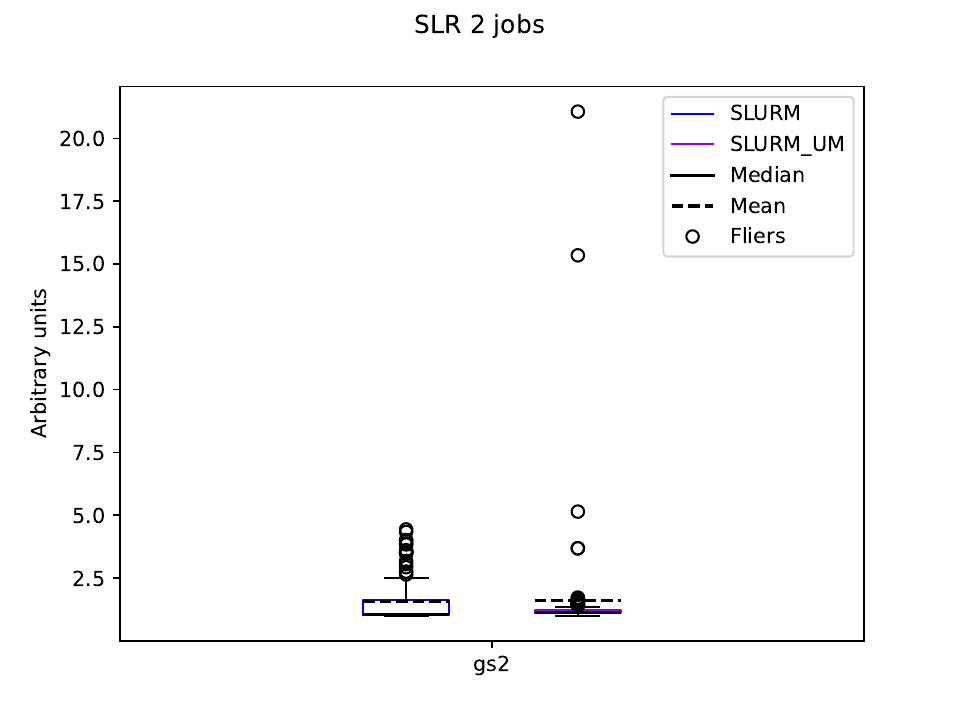}
    \includegraphics[width=0.4\textwidth]{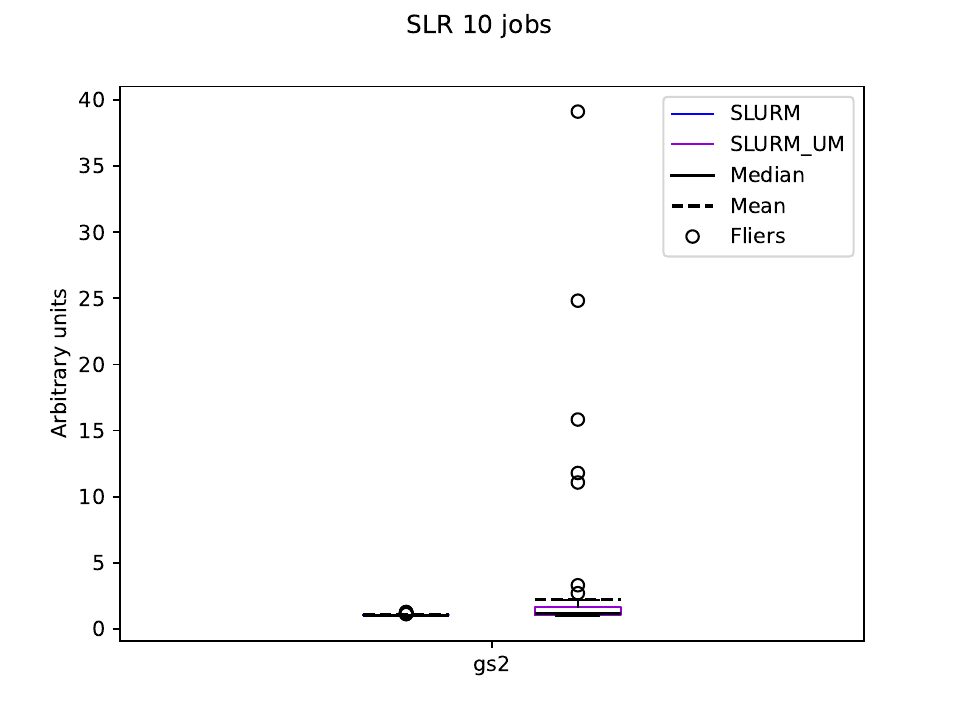}
    \caption{Boxplots showing the SLR for two jobs filling the queue (top) and 10 jobs filling the queue (bottom). The left (blue) box on the x-axis represents data collected from SLURM, and the right (purple) box represents data from the \acrshort{umbridge} SLURM backend.}
    \label{slrs um}
\end{figure}

\begin{figure*}[h]
\centering
\includegraphics[width=0.4\textwidth]{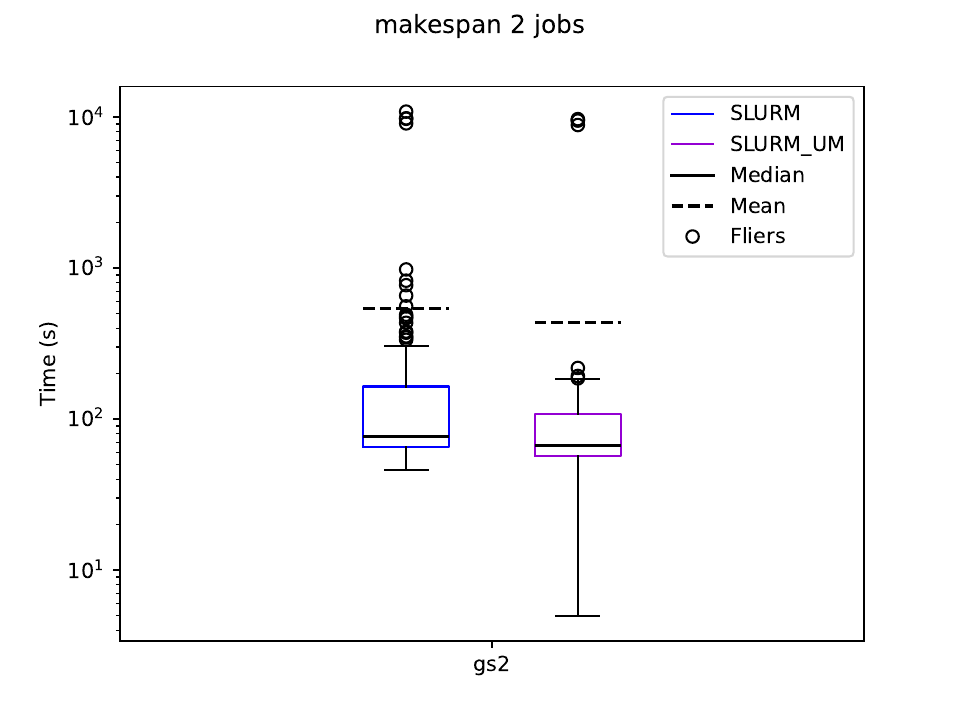} \includegraphics[width=0.4\textwidth]{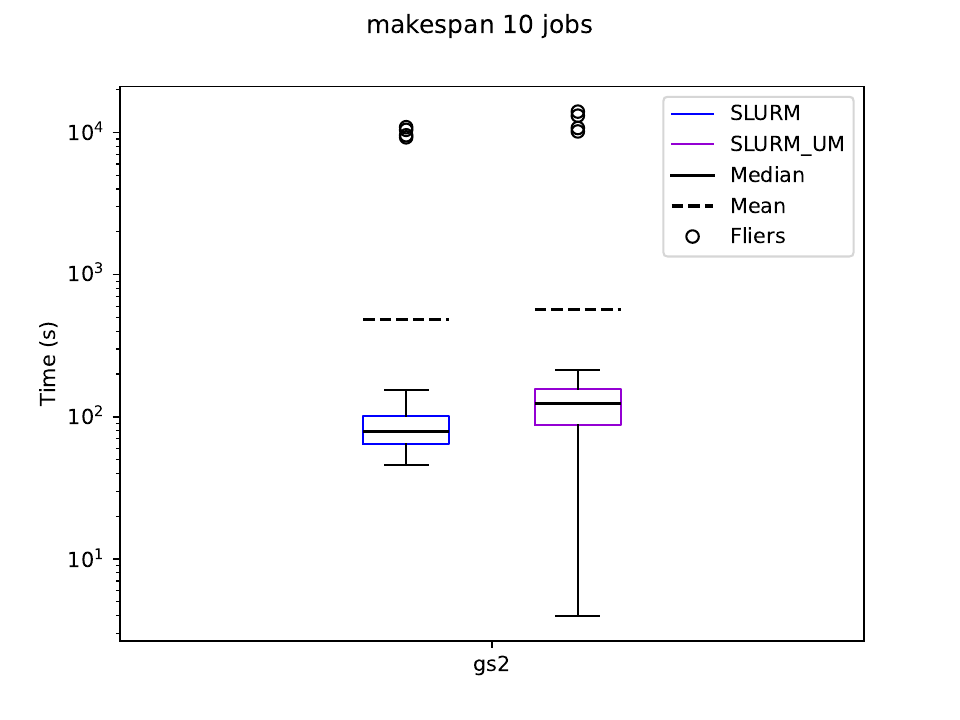}
\includegraphics[width=0.4\textwidth]{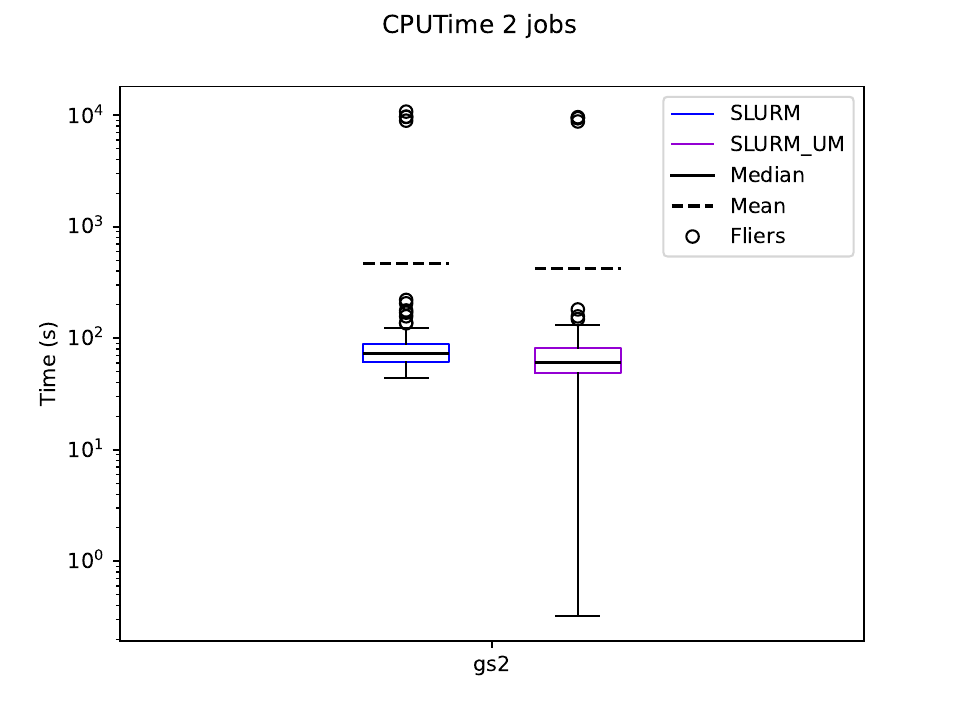}\includegraphics[width=0.4\textwidth]{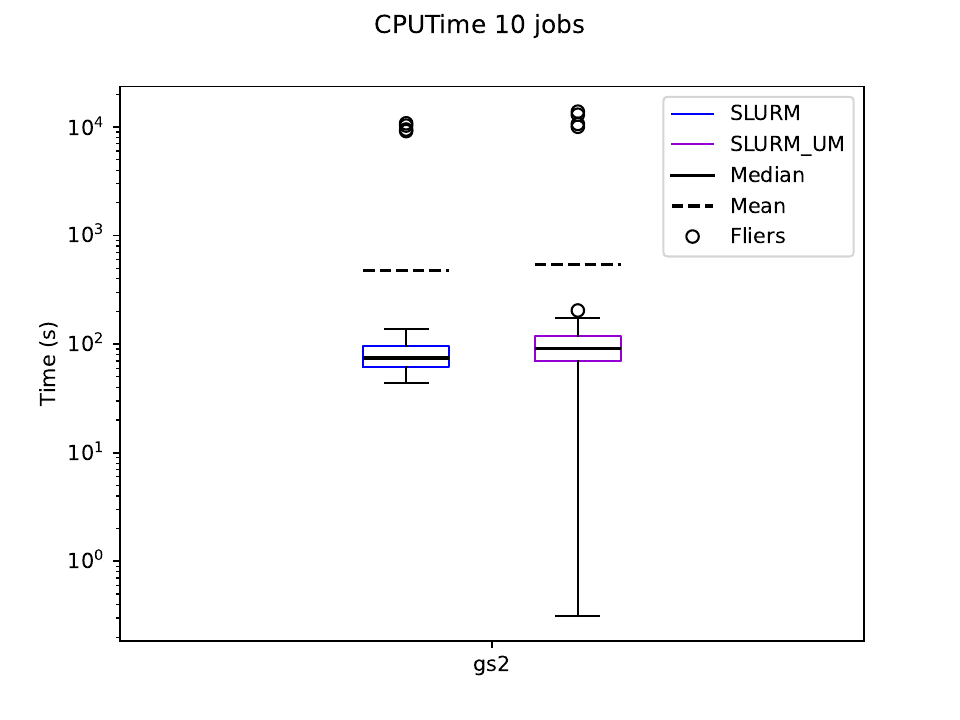}\\
\includegraphics[width=0.4\textwidth]{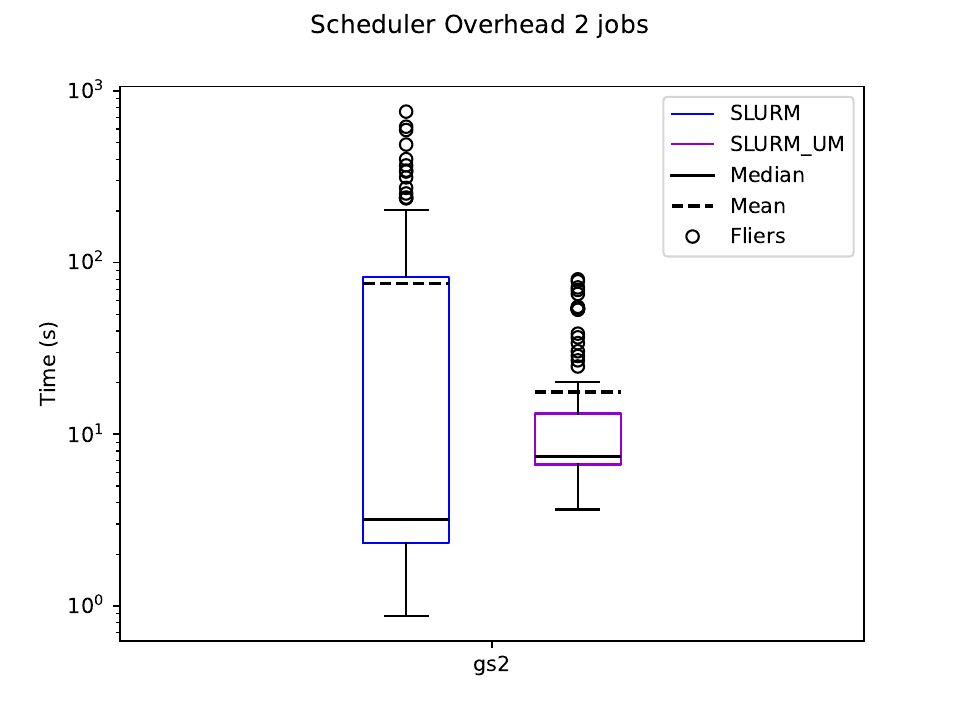}\includegraphics[width=0.4\textwidth]{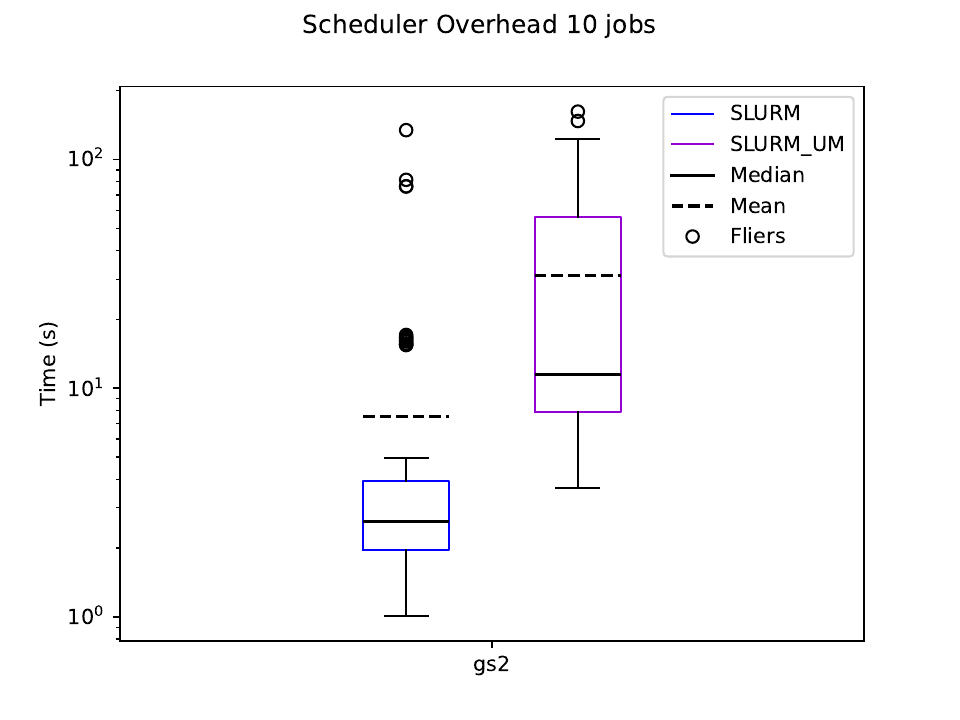}\\
\caption{Boxplots showing experimental results with 2 jobs (left column) and 10 jobs (right column) filling the queue. For the GS2 application on the x-axis, the left (blue) boxes represent data collected from SLURM, and the right (purple) boxes represent data from the \acrshort{umbridge} SLURM backend. The top row shows the makespan, the middle row the CPU time, and the bottom row shows the scheduler overhead, all measured in seconds.}
\label{2 jobs um}
\end{figure*}

The experimental outcomes comparing the na\"ive SLURM approach and the \acrshort{umbridge} SLURM backend (similar to Figure \ref{2 jobs} and \ref{slrs}) are plotted on Figure \ref{slrs um} and \ref{2 jobs um}. Only results from GS2 are presented in these plots.

\end{document}